\documentclass[reprint,twocolumn,showpacs,superscriptaddress,longbibliography,floatfix,aps,pra]{revtex4-2}

\usepackage{header}
\usepackage{comment}
\newcommand{\figref}[2]{Fig.~\hyperlink{#1-#2}{\ref{#1}#2}}
\newcommand{\figureref}[2]{Figure~\hyperlink{#1-#2}{\ref{#1}#2}}

\usepackage{xspace} 
\usepackage{ulem}

\newcommand{\up}{\ensuremath{\ket \uparrow}\xspace}
\newcommand{\down}{\ensuremath{\ket \downarrow}\xspace}

\newcommand{\updown}{\ensuremath{\dyad{\uparrow}{\downarrow}}\xspace}
\newcommand{\downup}{\ensuremath{\dyad{\downarrow}{\uparrow}}\xspace}
\newcommand{\upup}{\ensuremath{\dyad{\uparrow}{\uparrow}}\xspace}

\begin{document}
\title{Spectroscopic signatures of emergent elementary excitations in a kinetically constrained long-range interacting two-dimensional spin system}
\author{Tobias Kaltenmark}
\affiliation{Institut f\"ur Theoretische Physik, Universit\"at T\"ubingen, Auf der Morgenstelle 14, 72076 T\"ubingen, Germany}
\author{Chris Nill}
\affiliation{Institut f\"ur Theoretische Physik, Universit\"at T\"ubingen, Auf der Morgenstelle 14, 72076 T\"ubingen, Germany}
\affiliation{Institute for Applied Physics, University of Bonn, Wegelerstraße 8, 53115 Bonn, Germany}
\author{Christian Groß}
\affiliation{Physikalisches Institut, Universit\"at T\"ubingen, Auf der Morgenstelle 14, 72076 T\"ubingen, Germany}
\affiliation{Center for Integrated Quantum Science and Technology (IQST), Universit\"at T\"ubingen, Auf der Morgenstelle 14, 72076 T\"ubingen, Germany}
\author{Igor Lesanovsky}
\affiliation{Institut f\"ur Theoretische Physik, Universit\"at T\"ubingen, Auf der Morgenstelle 14, 72076 T\"ubingen, Germany}
\affiliation{Center for Integrated Quantum Science and Technology (IQST), Universit\"at T\"ubingen, Auf der Morgenstelle 14, 72076 T\"ubingen, Germany}
\affiliation{School of Physics and Astronomy and Centre for the Mathematics and Theoretical Physics of Quantum Non-Equilibrium Systems, The University of Nottingham, Nottingham, NG7 2RD, United Kingdom}

\begin{abstract}
    Lattice spin models featuring kinetic constraints constitute a paradigmatic setting for the investigation of glassiness and localization phenomena. The intricate dynamical behavior of these systems is a result of the dramatically reduced connectivity between many-body configurations. This truncation of transition pathways often leads to a fragmentation of the Hilbert space, yielding highly collective and therefore often slow dynamics. Moreover, this mechanism supports the formation of characteristic elementary excitations, which we investigate here theoretically in a two-dimensional Rydberg lattice gas. We explore their properties as a function of interaction strength and range, and illustrate how they can be experimentally probed  with a spectroscopic scheme. Here, we show that the transition rate to certain delocalized superposition states of elementary excitations displays collective many-body enhancement.
\end{abstract}

\maketitle

\section{Introduction}
The notion \textit{kinetic constraint} refers to the phenomenon that strong interactions or peculiar dynamical rules severely restrict the connectivity between many-body states in Hilbert space~\cite{Cancrini2009,Pavesic2025a}.
Such kinetically constrained systems are ubiquitous in physics. They naturally appear in the context of quantum circuits and quantum cellular automata, whose dynamics is governed by conditional state changes implemented by quantum gates~\cite{Gopalakrishnan2018}. Moreover, kinetic constraints play a crucial role in the onset of glassiness~\cite{Ritort2003, Chandler2010, Lesanovsky2013}, where the above-mentioned reduced connectivity between many-body states gives rise to extremely long relaxation times. Ultimately, they may even prevent thermalization by fragmenting the Hilbert space and give rise to the emergence of so-called quantum scars~\cite{Turner2018, Serbyn2021}.

A simple platform for studying the physics of kinetically constrained systems are strongly interacting Ising models~\cite{Islam2013}. In these models, constrained dynamics are implemented by exploiting large interaction-induced energy barriers, which prevent resonant (and therefore fast) transitions between certain spin configurations. In the past, this setting has been used to study a range of interesting phenomena, including confinement dynamics~\cite{Liu2019}, string-breaking~\cite{Verdel2020,De2024}, frozen states~\cite{Hart2022}, disorder-free (many-body) localization~\cite{Smith2016,Karpov2021}, and metastable phases~\cite{Rutkevich1999}.
Experimentally, such Ising models can be implemented with Rydberg atoms held in optical tweezer arrays~\cite{Browaeys2020}.
Their strong dipolar power-law interactions provide a straightforward route for implementing kinetic constraints~\cite{Valado2016}: strong interaction-induced energy shifts prevent the resonant laser excitation of many-body states containing many Rydberg excitations, thereby implementing the Rydberg blockade constraint \cite{Urban2009,Gaetan2009}.
In a many-body setting this mechanism realizes the so-called PXP-model, which has shown to host quantum scars~\cite{Chandran2023, Giudici2024}.
A further constraint is the so-called anti-blockade or facilitation~\cite{Ates2007, Amthor2010, Schempp2014, Urvoy2015, Valado2016, Marcuzzi2017, Letscher2017, Ostmann2019, Magoni2021a, Klocke2021, Liu2022, Zhang2024, Brady2024}. Due to the appearance of strong mechanical forces \cite{Faoro2016} --- manifesting in spin-phonon coupling~\cite{Magoni2021b, Magoni2024, Brady2025, Lienhard2025} --- such a scenario is challenging to implement experimentally, although recent progress has been made in one dimension~\cite{Zhao2025}.
In the facilitation regime, the Rydberg excitation of an atom located next to an already excited atom becomes resonantly enhanced.
This manifests in an avalanche-like dynamics, nucleation dynamics~\cite{Osterholz2025} and, in conjunction with dissipative decay processes, enables the study of non-equilibrium universality~\cite{Gutierrez2017, Helmrich2020}.

In this work, we investigate elementary excitations in a two-dimensional square lattice from which Rydberg atoms are laser-excited under the facilitation constraint. We show how this constraint, as well as the residual power-law tail of the interactions, determines the energetic and kinetic properties of these excitations. Moreover, we develop an effective theory for one excitation class consisting of chains of Rydberg atoms, enabling us to investigate their dynamics on large lattices. Finally, we discuss how to experimentally probe these elementary excitations via a modulated laser. Here, we also identify spectroscopic signatures of collective many-body enhancement.


\section{System}
\begin{figure}
    \centering
    \includegraphics[width=\linewidth]{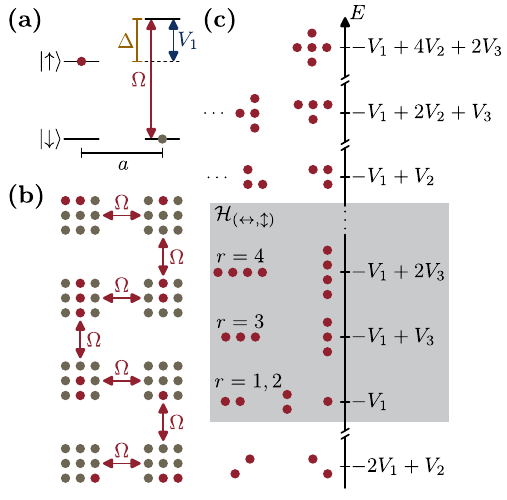}
    \caption{\textbf{Elementary excitations in a facilitated two-dimensional Rydberg square lattice.}
    \textbf{(a)}~Level scheme of two atoms separated by one lattice constant $a$. The atoms are modelled by two states, a ground state $\down$ (beige) and a Rydberg state $\up$ (red). Both states are coupled by a laser with Rabi frequency $\Omega$ and detuning $\Delta$. The latter is chosen to compensate the nearest-neighboring Rydberg interaction, i.e.\,, $V_1+\Delta=0$, which is the so-called facilitation condition.
    \textbf{(b)}~Some near-resonant processes in a 3$\times$3-lattice, assuming that $\Omega$ is much larger than the next-next-nearest-neighbor interaction $V_3$.
    \textbf{(c)}~Structure and energy of some elementary excitations. Here, $V_2$ and $V_3$ are the interaction energies between next-nearest and next-next-nearest neighbors, respectively.
    The grey box marks a subspace $\mathcal H_{(\leftrightarrow,\updownarrow)}$, where elementary excitations are formed by chains of Rydberg atoms.
    These carry energy $r\Delta$ from their $r$ Rydberg excitations and energy $(r-1)V_1$ from the $r-1$ nearest-neighbor pairs. Under facilitation condition, this implies a constant energy shift $-V_1$ for these elementary excitations.
    }
    \label{fig:Overview}
\end{figure}

We consider a two-dimensional square lattice with dimensions $N_x,N_y$, and $N=N_x\times N_y$ sites. Each lattice site is occupied by a single atom, which is modelled as a two-level system, forming a fictitious spin $1/2$ particle, see \figref{fig:Overview}{(a)}. A laser with Rabi frequency $\Omega$ and detuning $\Delta$ couples the electronic ground state $\down$ of an atom to the Rydberg state $\up$, and atoms in the Rydberg state interact via the Van der Waals interaction. The Hamiltonian that models this system is given by 
\begin{equation}\label{eq:exact Hamiltonian}
    H=\sum_{\bm j=(1,1)}^{(N_x,N_y)}\left(\frac{\Omega}{2} \sigma_{\bm j}^{(x)}+ \Delta n_{\bm j}\right)+ \frac12 \sum_{\bm j\neq\bm k}\frac{V_1}{|\bm j-\bm k|^6}n_{\bm j}n_{\bm k}.
\end{equation}
The first term describes electronic transitions within the atoms that are induced by the laser: $\bm j=(j_x,j_y)$ refers to the lattice site, $\sigma_{\bm j}^{(x)}=\updown_{\bm j}+\downup_{\bm j}=\sigma_{\bm j}^++\sigma_{\bm j}^-$ is the Pauli $x$-matrix and $n_{\bm j}=\upup_{\bm j}$ is the projector onto the Rydberg state of the $\bm j$-th atom. The second term models the interaction between excited Rydberg atoms, where the coefficient $V_1=C_6/a^6$ is the interaction strength between Rydberg atoms occupying nearest-neighboring lattice sites. It depends on the dispersion coefficient $C_6$ and the lattice spacing $a$. Given that the interaction rapidly decays with distance, we only consider interactions between nearest neighbors, $V_1$, next-nearest neighbors, $V_2=V_1/2^3$, and next-next-nearest neighbors, $V_3=V_1/2^6$. We assume furthermore that Rydberg atoms are excited under facilitation condition, which means that the laser detuning is cancelled by the nearest-neighbor interaction, $\Delta+V_1=0$. When at the same time the Rabi frequency is sufficiently small, $\Omega/2\ll V_1$, the excitation dynamics is effectively kinetically constrained: an atom can only be resonantly excited, when it is located next to exactly one Rydberg atom. Dynamically, this leads to a highly correlated formation of Rydberg excitations. An example of this is shown in \figref{fig:Overview}{(b)} for a $3\times 3$ lattice.


\section{Elementary excitations.---}
To illuminate the nature of these correlated elementary excitations, we consider the spectrum of Hamiltonian~\eqref{eq:exact Hamiltonian} in the limit $\Omega=0$.
Under this condition, the eigenstates are product states in the classical spin basis $\{\up,\down\}$.
In \figref{fig:Overview}{(c)}, we sketch a few of those states along with their energy. In this work, we focus on the states highlighted in the grey box,
whose energy separation is multiples of the next-next-nearest neighbor interaction $V_3$. Assuming that $\Omega/2\gg V_3$, these states form a near-resonant subspace $\mathcal H_{(\leftrightarrow,\updownarrow)}$ in which many-body configurations are strongly coupled by the laser.
As can be seen, Rydberg excitations in $H_{(\leftrightarrow,\updownarrow)}$ are forced to form consecutive chains, whose overall number is conserved: they can neither vanish, nor can two or more chains merge.
This effect has been studied in detail in one-dimensional lattices~\cite{Marcuzzi2017,Magoni2021a,Magoni2021b,Magoni2024}.
In two dimensions, the situation is more involved as these chains can grow in vertical and horizontal directions. Nevertheless, excitations can explore the entire lattice and are therefore mobile, see \figref{fig:Overview}{(b)}. 
In contrast, there are also immobile excitations which are formed by configurations in which Rydberg atoms populate next-nearest neighboring lattice sites, where they interact with strength $V_2$. As an example, consider the triangular configuration with energy $-V_1+V_2$, shown in \figref{fig:Overview}{(c)}. Assuming that $\Omega/2\ll V_2$, the Rydberg atoms in the diagonal lattice sites cannot be resonantly removed. Nevertheless, further Rydberg atoms can be added resonantly.
Consequently, the overall dynamics reduce to that of pinned one-dimensional chains that can grow and shrink, but are unable to explore the full lattice.

The dynamics of the mobile excitations are described by the Hamiltonian 
\begin{align}\label{eq:reduced-Hamiltonian}
    H_{(\leftrightarrow, \updownarrow)}=&\frac{\Omega}{2}\mathbf T+V_3\sum_{r=3}^N(r-2)\mathbf P_r-V_1.
\end{align}
Here, $\mathbf T$ is a kinetic term, that gives rise to expansion and shrinking of Rydberg chains. It dynamically connects Rydberg chains that differ by a single Rydberg atom, as shown in \figref{fig:Overview}{(b)}.
The second term assigns an energy to each chain, which is proportional to its length $r$ and is given by the number of next-next-nearest neighbored Rydberg atom pairs. Here, the operator $\mathbf P_{r}$ projects onto all vertical and horizontal chains with length $r$. The last, constant, contribution results from the energetic "cost" $r\Delta$ of exciting $r$ Rydberg atoms, which is partially compensated by $(r-1)V_1$ nearest-neighbor interactions. With the facilitation condition, this results in a constant energy shift of $-V_1$ for all Rydberg chains. The derivation of Hamiltonian (\ref{eq:reduced-Hamiltonian}) is shown in the End Matter, where we also provide the explicit forms of $\mathbf T$ and $\mathbf P_{r}$. Furthermore, we show that the dimension of the subspace $\mathcal H_{(\leftrightarrow,\updownarrow)}$ grows only polynomially in the system size, which dramatically simplifies numerical simulations. 

\begin{figure}
    \centering
    \includegraphics[width=\linewidth]{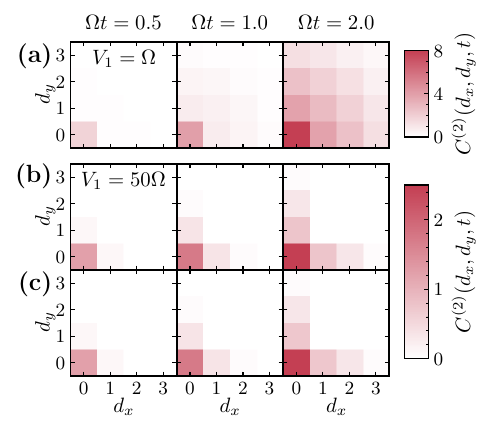}
    \caption{\textbf{Emergence of the kinetic constraint.} Correlation function $C^{(x)}(d_x,d_y,t)$ in a $4\times 4$ lattice with an initial single excitation positioned at site $(2,2)$. 
    \textbf{(a,b)}~Data obtained by evolving the system under the full Hamiltonian \eqref{eq:exact Hamiltonian}, for weak interaction $V_1=\Omega$ and strong interaction $V_1=50\Omega$, respectively. In the latter case, a single Rydberg excitation initiates the growth of chains of Rydberg excitations along the rows and columns of the lattice [see Fig. \ref{fig:Overview}{(b)}].
    \textbf{(c)}~Correlations calculated from the reduced model \eqref{eq:reduced-Hamiltonian}, for $V_1=50\Omega$. Rows~\textbf{(b,c)} are in excellent agreement, which demonstrates that the reduced model description becomes appropriate when the interaction strength is sufficiently large. The snapshots are taken at times $\Omega t=0.5,1,2$.}
    \label{fig:Correlations}
\end{figure}

In the following we benchmark the reduced model $H_{(\leftrightarrow, \updownarrow)}$ against the full Hamiltonian (\ref{eq:exact Hamiltonian}). To this end we time-evolve an initial state with a  single Rydberg excitation at site $(2,2)$ in a $4\times4$ lattice. For both models, we compute the integrated density-density correlations $C^{(2)}(\bm d,t)=\sum_{\bm j}\ev{n_{\bm j} n_{\bm j+\bm d}}_t$. Here, $\ev{n_{\bm j} n_{\bm j+\bm d}}_t$ quantifies the probability of an atom at site $\bm j$ to be excited simultaneously with another atom at a distance $\bm d=(d_x,d_y)$. 
In \figref{fig:Correlations}{(a)}, we show the density-density correlations in case of weak interaction, i.e. $V_1=\Omega$, computed for the full Hamiltonian (\ref{eq:exact Hamiltonian}). Here, fluctuations are strong enough to overcome diagonal energy contributions $V_2$. This is reflected in an isotropic build-up of correlations over time.
However, when $V_2\gg\Omega/2$, transitions including $V_2$ are effectively forbidden and correlations are only formed between atoms that are located on the same row or column, see \figref{fig:Correlations}{(b)}. Here only states in the subspace $\mathcal H_{(\leftrightarrow,\updownarrow)}$, i.e. Rydberg chains, are populated and the dynamics is faithfully described by the reduced model~(\ref{eq:reduced-Hamiltonian}). This can be seen in \figref{fig:Correlations}{(c)}, which shows data that are effectively indistinguishable from the full simulation.


\begin{figure*}
    \centering
    \includegraphics[width=\linewidth]{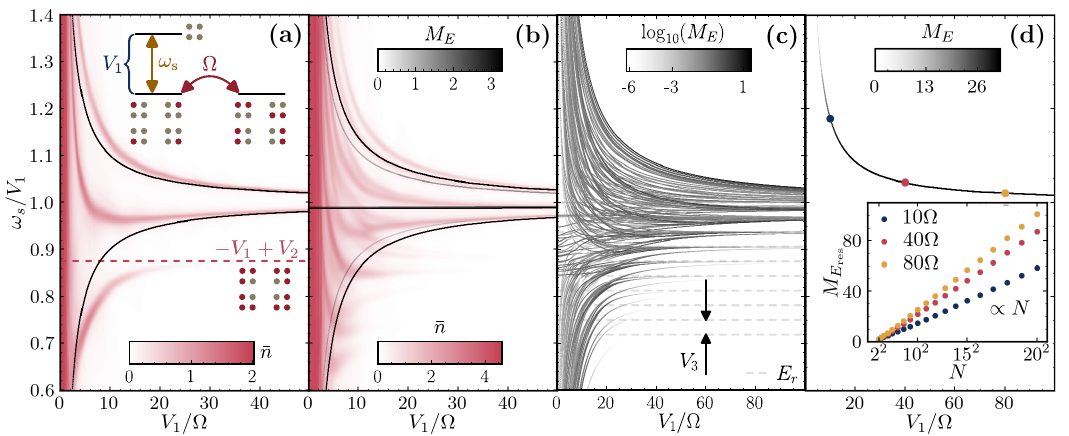}
    \caption{\textbf{Spectroscopy and collective enhancement.} \textbf{(a,b)} Excitation spectrum for a $2\times2$ and $3\times3$ lattice as a function of the interaction strength $V_1$ and modulation frequency $\omega_\text s$. In red we show the time-averaged number of Rydberg atoms $\bar n$, which is calculated considering the full dynamics under the Hamiltonian (\ref{eq:Ham spectroscopy}), with $\Omega_\text s=0.2\,\Omega$. To evaluate $\bar n$ we discretize the integral with sampling points $\Omega t\in\{1,2,\cdots,30\}$. The grey scale shows the transition matrix element $M_E$ at the expected resonance energy, computed with the spectrum of the reduced model Hamiltonian (\ref{eq:reduced-Hamiltonian}). Note, that the matrix elements in a $2\times2$ lattice are constant ($M_E=2$), in contrast to the matrix elements in a $3\times3$ lattice. This is a result of the next-next-nearest neighbor interaction $V_3$, which does not contribute for a $2\times2$ lattice. The inset of panel \textbf{(a)} illustrates the idea behind the spectroscopic scheme: the Rabi frequency modulation at frequency $\omega_\text s$ probes excitations, which are formed by (kinetically constrained) many-body states that are near-resonantly coupled by the laser (Rabi frequency $\Omega$). For intermediate interaction strength we observe excitations of many-body configurations that are not contained in the subspace $\mathcal H_{(\leftrightarrow,\updownarrow)}$, e.g. triangular patterns of Rydberg atoms with energy $-V_1+V_2$. With increasing interaction strength only excitations contained in $\mathcal H_{(\leftrightarrow,\updownarrow)}$ are excited. \textbf{(c)} Spectrum of an $11\times11$ lattice, calculated with the reduced model (\ref{eq:reduced-Hamiltonian}). Here, we used a logarithmic scale $\log_{10}(M_E)$ to increase the contrast of the spectral lines. The grey dashed lines correspond to the energies $-V_1+(r-2)V_3$ of Rydberg chains with length $r\ge2$, see End Matter. \textbf{(d)} Collectively enhanced spectral line in the $11\times11$ lattice. The inset depicts the transition matrix element $M_{E_\text{res}}$ evaluated with the corresponding eigenstate at energy $E_\text{res}$ for different lattice sizes $N=N_x\times N_y$ with $N_x=N_y\in\{2,3,\cdots,20\}$.}
    \label{fig:spectroscopy}
\end{figure*}
\section{Spectroscopy}In the following, we propose a spectroscopic scheme to experimentally probe elementary excitations 
and to measure their energies. To this end, we initialize the system in a state without Rydberg atoms, i.e. $\ket 0=\down^{\otimes N_xN_y}$, which is the natural starting point of any experiment.
When $V_1\gg\Omega/2$, this state is kinetically disconnected from the remaining Hilbert space and merely acquires a second order light shift.
On the contrary, many-body configurations that host Rydberg atoms, in particularly those contained in $\mathcal H_{(\leftrightarrow,\updownarrow)}$ are generally near-resonantly coupled by the laser. 
This leads to the formation of the mobile elementary excitations, i.e.\,, many-body eigenstates that are formed from superpositions of the classical configurations shown in \figref{fig:Overview}{(c)}.
To probe these, we periodically modulate the Rabi frequency around a central value $\Omega/2$ with frequency $\omega_\text s$ and small amplitude $\Omega_\text s\ll\Omega/2$.
Thus, the the many-body Hamiltonian is modified according to
\begin{equation}\label{eq:Ham spectroscopy}
    H'(t)=H+\frac{\Omega_\text s}{2}\cos(\omega_\text s t)\sum_{\bm j}\sigma_{\bm j}^{(x)}.
\end{equation}
When the modulation frequency $\omega_\text s$ matches the eigenenergy of an elementary excitation, Rydberg atoms can be coherently excited, see inset of \figref{fig:spectroscopy}{(a)}.
We consider in the following the time-averaged Rydberg density, $\bar n=\frac1\tau\int_0^\tau\mathrm dt\langle \sum_{\bm j} n_{\bm j}\rangle_t$, where $\tau$ is the length of the averaging time window. Note, that the integration time $\tau$ has to be long enough in order to reach sufficient spectral resolution, see~\figref{fig:time-resolved spectroscopy}{} in the End Matter. In \figref{fig:spectroscopy}{(a,b)} we show $\bar n$ as a function of $\omega_s$ and $V_1$ for a $2\times2$ and $3\times 3$ lattice. For strong nearest-neighbor interaction, i.e. when the facilitation constraint is present, one observes sharp spectral features at specific modulation frequencies. To show that these spectral lines correspond to elementary excitations described by reduced model Hamiltonian (\ref{eq:reduced-Hamiltonian}) we compute it's eigenvalues $E$ and eigenstates $\ket{E}$. From these, we calculate the expected resonance positions $\omega_\text s=|E|$ and show them as grey/black curves in \figref{fig:spectroscopy}{(a,b)}. Indeed, when $V_1$ is sufficiently large, i.e. when the facilitation constraint is present and the reduced model is valid, there is excellent agreement with the data computed from the full Hamiltonian (\ref{eq:Ham spectroscopy}).
The transition rate, i.e. the strength of the spectral lines obtained in the reduced model, can be estimated by Fermi's golden rule. It is proportional to the transition matrix element $M_E=|\matrixel{E}{\sum_{\bm j}\sigma_{\bm j}^{(x)}}{0}|^2=|\matrixel{E}{\sum_{\bm j}\sigma_{\bm j}^{+}}{0}|^2$, which we plot in grey scale in \figref{fig:spectroscopy}. Note, that for small nearest-neighbor interactions $V_1$, we also observe signatures of  excitations, which are not captured by the reduced model~\eqref{eq:reduced-Hamiltonian}. For example, in the $2\times2$ lattice we observe one spectral feature at $\omega_{\text s}\approx |-V_1+V_2|$, which belongs to excitations in which Rydberg atoms form a triangular pattern, see also \figref{fig:Overview}{(c)}.
As the Rydberg interaction grows, exciting those states becomes increasingly hard. This is because these states can only be accessed from the initial state $\ket 0$, via the excitation of multiple Rydberg atoms through a sequence of off-resonant processes.

As a final study, we analyze the excitation spectrum of an $11\times 11$ lattice using the reduced model Hamiltonian (\ref{eq:reduced-Hamiltonian}). We calculate the resonance positions from the eigenenergies and show the corresponding transition matrix element $M_E$ in \figref{fig:spectroscopy}{(c)}. We observe several transition lines. For large $V_1$, they are located at $\omega_{\text s}=V_1-V_3(r-2)$ with $r\in\{2,\cdots 11\}$, which indicates that they correspond to excitations that are mainly formed by chains of Rydberg atoms of length $r$, see \figref{fig:Overview}{(c)}. Note, that we use a logarithmic color scale as the strength of the lines varies considerably. Plotting the data on a linear scale, as in \figref{fig:spectroscopy}{(d)}, reveals that there is one dominant (collectively enhanced) line whose strength grows proportional to the size $N$ of the lattice, see inset. This is in fact the strongest possible scaling with system size, which can be seen by writing the matrix element as $M_E=N|\braket{E}{s}|^2\le N$. Here, $\ket s=\frac{1}{\sqrt N}\sum_{\bm j}\sigma_{\bm j}^+\ket 0$ is a spin wave state, which corresponds to a single Rydberg excitation delocalized over the entire lattice. This shows that the elementary excitation that corresponds to the collectively enhanced line is predominantly formed by many-body states that contain short Rydberg chains and single excitations.  



\section{Conclusions and outlook} A two-dimensional long-range interacting Rydberg gas under facilitation conditions features a host of elementary excitations. We have developed an effective theory for a particular class of mobile excitations, which emerge when nearest-neighbor and next-nearest-neighbor interaction are sufficiently strong. Their energies can be spectroscopically determined by applying an appropriately modulated excitation laser and a subsequent measurement of the Rydberg atom number. In the future it would be interesting to investigate how the properties of the elementary excitations are affected when the condition of strong next-nearest-neighbor interaction is systematically relaxed. If solely nearest-neighbor interactions are present, the physics reduces to that of a two-dimensional transverse field Ising model, which already displays intriguing collective properties, such as metastability \cite{Pavesic2025b} and quasi-particle confinement \cite{Pavesic2025a}. Therefore, it would be interesting to explore many-body relaxation processes and their dependence on the facilitation constraint as well as the residual long-range interactions.
Moreover, a quantitative characterization of transport properties, such as diffusion constants, as well as long-time relaxation dynamics represent a particularly promising avenue for future research.
Computationally, these will be formidable tasks. However, quantum simulators based on trapped Rydberg atoms may ultimately offer such capability. They could thus yield novel insights into relaxation pathways and timescales, which may allow to identify glassy dynamics. Currently though, the required long observation times still pose a technical challenge.


\acknowledgments
\textit{Acknowledgements.---}
The analysis code and data underlying the results presented in this paper are freely accessible in the open-access Zenodo database, see~\cite{Zenodo}.
We acknowledge funding from the Deutsche Forschungsgemeinschaft (DFG, German Research Foundation) through the Research Unit FOR 5413/1, Grant No. 465199066 and 465199066, and through the Research Unit FOR 5522/1, Grant No. 499180199. 
We acknowledge support from the Leverhulme Trust (Grant No. RPG-2024-112) and the European Union under the Horizon Europe program HORIZON-CL4-2022-QUANTUM-02-SGA via the project 101113690 (PASQuanS2.1).
We also acknowledge funding through JST-DFG 2024: Japanese-German Joint Call for Proposals on “Quantum Technologies” (Japan-JST-DFG-ASPIRE 2024) under DFG Grant No. 554561799.  This work is also supported by the ERC grant OPEN-2QS (Grant No. 101164443). 
CG acknowledges support by the Alfried Krupp von Bohlen and Halbach foundation.
\appendix

\section{Reduced Hilbert space and coordinates}
A chain of Rydberg atoms, i.e. a many-body state in the reduced Hilbert space $\mathcal H_{(\leftrightarrow,\updownarrow)}$, can be uniquely specified by four coordinates $(c_x,r_x,c_y,r_y)$. The relative coordinates $(r_x,r_y)$ encode the length of the chain in the horizontal and vertical directions, respectively. For open boundary conditions, $r_k$ is an integer in the range 1,\,\dots,\,$N_k$ for $k \in \{x,y\}$. Note, that vertical chains ($r_y \ge 2$) have a width of one atom in the horizontal direction ($r_x = 1$), and vice versa for horizontal chains. Accordingly, we impose
\begin{align}
    r_k \ge 2 \Longrightarrow r_l = 1 \qquad \text{for } k \neq l.
\end{align}

The remaining coordinates $\mathbf c = (c_x,c_y)$ denote the center of the chain on the two-dimensional lattice. For chains of odd length, this center coincides with a lattice site, whereas for even chain lengths it lies halfway between two neighboring lattice sites. We therefore choose centers coinciding with lattice sites as odd integers $c_k$, and centers located between two sites as even integers. The center coordinates map to lattice sites via $\bm j = \bigl(\tfrac{c_x+1}{2},\tfrac{c_y+1}{2}\bigr)$, so that $c_k$ takes integer values in the range $1,\dots,2N_k-1$ for $k=x,y$. Since each center coordinate pair $(c_k,r_k)$ is chosen such that either both coordinates are even or odd, the sum $c_k+r_k$ and difference $c_k-r_k$ are even integers. Therefore, they refer to positions between two lattice sites and give for horizontal/vertical Rydberg chains the right/upper ($c_k+r_k$) and left/lower ($c_k-r_k$) boundary, respectively. With this, we enforce the boundary conditions that Rydberg chains are not allowed to exceed the lattice, and require
\begin{align}\label{eq:boundary condition lattice}
    0 \le c_k - r_k, && 2N_k \ge c_k + r_k, &&\text{for } k\in\{x,y\}.
\end{align}
Here, we used 0 for the left lattice boundary and $2N_k$ for the right boundary.

Combining all constraints, the coordinates for horizontal chains (fixed $r_x \ge 2$) are given by $c_x\in\{r_x, r_x+2, \dots, 2N_x - r_x\}$, $r_y = 1$, and $c_y \in \{1,3,\dots,2N_y-1\}$. Hence, there are $(N_x - r_x + 1)\cdot 1 \cdot N_y$ distinct horizontal chains of length $r_x$. Similarly, there are $N_x \cdot 1 \cdot (N_y - r_y + 1)$ vertical chains of fixed length $r_y \ge 2$. The dimension of the reduced space $\mathcal H_{(\leftrightarrow,\updownarrow)}$ is therefore obtained by summing over all admissible chain lengths $r_x \in \{2,\dots,N_x\}$ and $r_y \in \{2,\dots,N_y\}$, and adding the $N_x \times N_y$ configurations with a single Rydberg excitation ($r_x = r_y = 1$), yielding
\begin{align}
    \dim\bigl(\mathcal H_{(\leftrightarrow,\updownarrow)}\bigr)
    = \frac{N_x N_y}{2}\,(N_x + N_y).
\end{align}
Thus, the dimension of the reduced space scales polynomially with the system size.

In the following we construct the projection of the Hamiltonian \eqref{eq:exact Hamiltonian} onto the subspace $\mathcal H_{(\leftrightarrow,\updownarrow)}$, which results in the Hamiltonian $H_{(\leftrightarrow,\updownarrow)}$, Eq. \eqref{eq:reduced-Hamiltonian}. The first term in the Hamiltonian \eqref{eq:exact Hamiltonian} induces single spin flips (electronic state changes due to the laser). Due to the facilitation constraint these flips lead in $\mathcal H_{(\leftrightarrow,\updownarrow)}$ to processes that expand and shrink the size of a chain of Rydberg atoms by one atom. Therefore, horizontal chains can change their size in $x$-direction via four different processes: expansion or shrinking at the right boundary via the operators $\dyad{c_x+1}{c_x}\otimes\dyad{r_x+1}{r_x}+\dyad{c_x}{c_x+1}\otimes\dyad{r_x}{r_x+1}$, and the same processes at the left boundary, represented by $\dyad{c_x}{c_x+1}\otimes\dyad{r_x+1}{r_x}+\dyad{c_x+1}{c_x}\otimes\dyad{r_x}{r_x+1}$. Note that a change in the relative coordinate also implies a change of the center of mass position. All these processes, which leave the position and width in the $y$-coordinate invariant, are represented by the kinetic energy operator
\begin{widetext}
\begin{align}
    T_x&=\sum_{c_x=1}^{2N_x-2}\sum_{c_y=1}^{2N_y-1}\sum_{r_x=1}^{N_x-1}\sum_{r_y=1}^{N_y}(\dyad{c_x+1}{c_x}+\text{h.c.})\otimes(\dyad{r_x+1}{r_x}+\text{h.c.})\otimes\dyad{c_y,1}{c_y,1}.
\end{align}
\end{widetext}
Note, that here $1\leq c_x \leq 2N_x-2$ and $1\leq r_x\leq N_x-1$ which is due to the boundary conditions. Through similar reasoning, where $c_x,r_x$ and $c_y,r_y$ interchange roles, we obtain the corresponding expression for the kinetic energy vertical chains $T_y$. Combining both yields the total kinetic energy operator $\mathbf T=T_x+T_y$, see Eq.~\eqref{eq:reduced-Hamiltonian}. 

The remaining part of Hamiltonian~\eqref{eq:reduced-Hamiltonian} is diagonal and assigns an energy to each basis state. This energy only depends on the number of excited atoms, and the number of interacting Rydberg pairs. Therefore, for a chain with length $r=\max{(r_x,r_y)}$, the energy is given by
\begin{align}
    E_r=&r\Delta+\Theta(r-1)V_1(r-1)+\Theta(r-2)V_3(r-2) \nonumber\\
    E_r\stackrel{\text{fac.}}=&-V_1+\Theta(r-2)(r-2)V_3. \label{eq:energy Rydberg chain}
\end{align}
Here, $\Theta(x)$ is the Heaviside step function, which is equal to zero for $x\le0$ and one for $x>0$. Note, that there is no contribution containing $V_2$, because Rydberg chains do not include pairs of excited next-nearest-neighboring atoms. Finally, we define the projection operators 
\begin{align}
    P_{(r_x,r_y)}&=\sum_{c_x,c_y=1}^{2N_x-1,2N_y-1}\dyad{c_x,r_x,c_y,r_y}{c_x,r_x,c_y,r_y} \nonumber\\
    \mathbf P_r&=P_{(r,1)}+P_{(1,r)},
\end{align}
which project all vertical and horizontal chains of length $r$ onto itself. Together with the kinetic energy operator this yields Hamiltonian \eqref{eq:reduced-Hamiltonian} of the main text.

Lastly, we provide an expression for the number operators $n_{\bm j}$ in the reduced space $\mathcal H_{(\leftrightarrow,\updownarrow)}$, which is necessary for calculating the correlations in the main part. 
The number operator $n_{\bm j}$ projects the atom at site $\bm j$ to the excited state and leaves the state of the remaining atoms invariant.
To conserve this structure in the reduced space, we use the coordinates of a Rydberg chain $\ket{c_x,r_x,c_y,r_y}$, to determine if it includes a Rydberg atom at site $\bm j$. 
For this, we first convert the site $\bm j$ to the center coordinates with the transformation above, yielding $\bm c_{\bm j}=(2j_x-1,2j_y-1)$. From the derivation of $\dim(\mathcal H_{(\leftrightarrow,\updownarrow)})$, we use that the boundaries of a Rydberg chain are given by $(c_k-r_k)$ and $(c_k+r_k)$, for $k=x,y$. A Rydberg chain $\ket{c_x,r_x,c_y,r_y}$, then includes an excitation at lattice site $\bm j$, if $\bm c_{\bm j}$ is within the boundaries of the Rydberg chain, i.e.
\begin{align}\label{eq:number operator condition reduced space}
    \begin{pmatrix}
            c_x-r_x\\
            c_y-r_y
        \end{pmatrix}
        &\le
        \begin{pmatrix}
            2j_x-1\\
            2j_y-1
        \end{pmatrix}
        \le
        \begin{pmatrix}
            c_x+r_x\\
            c_y+r_y
        \end{pmatrix}.
\end{align}
The number operator $n_{\bm j}$ in the reduced space, projects onto all states satisfying this condition, i.e.\,, all states in $\mathcal N_{\bm j}=\{\ket{c_x,r_x,c_y,r_y}\in\mathcal H_{(\leftrightarrow,\updownarrow)}|\text{Eq. }\eqref{eq:number operator condition reduced space}\text{ is satisfied}\}$, and reads
\begin{align}
    n_{\bm j}=\sum_{\ket{c_x,r_x,c_y,r_y}\in\mathcal N_{\bm j}}\dyad{c_x,r_x,c_y,r_y}.
\end{align}

\begin{figure}
    \centering
    \includegraphics[width=\linewidth]{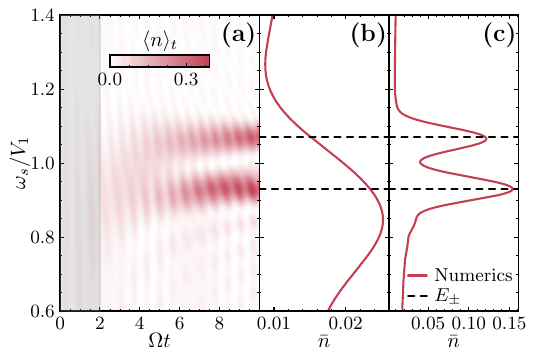}
    \caption{\textbf{Time-resolved spectroscopy signal of two facilitated atoms.}
    \textbf{(a)}~Time evolution of the Rydberg density $\langle n\rangle_t$ under a modulated laser of two facilitated atoms for varying modulation frequency $\omega_\text s$ and $\Omega_\text s=0.2\,\Omega$, and $V_1=10\,\Omega$.
    \textbf{(b)}~Time-averaged Rydberg density $\bar n$ in the time window, indicated by the grey shading in panel \textbf{(a)}. The average $\bar n$ is evaluated for $\tau=2/\Omega$, using $200$ equidistant sampling points of the integral. The red line depicts $\bar n(\omega_\text s)$ and the dashed black lines indicate the theoretical eigenvalues at $\omega_\text s=|E_\pm|$, Eq. \eqref{eq:eigenenergy two facilitated atoms}.
    \textbf{(c)}~Time-averaged Rydberg density $\bar n$ for the full time window ($\tau=10/\Omega$). The integral is evaluated with $1000$ equidistant sampling points. We observe that the spectral lines become only visible for sufficiently long coherent evolution times.}
    \label{fig:time-resolved spectroscopy}
\end{figure}
\section{Spectroscopy and coherent evolution.}
In Fig.~\ref{fig:spectroscopy} we show the time-averaged density of Rydberg atoms, $\bar n$, as a function of the laser frequency $\omega_\text s$.
In this figure a number of spectral lines are clearly visible.
In the following we discuss the impact of the overall coherent evolution time on the observability of these features. To make the point, we consider a system of two atoms. Under facilitation conditions, ($\Delta+V_1= 0$ and $V_1\gg \Omega/2$), their dynamics is described by Hamiltonian \eqref{eq:exact Hamiltonian} and the subspace $\mathcal H_{(\leftrightarrow,\updownarrow)}$ is spanned by the states $\ket{\uparrow\downarrow},\ket{\downarrow\uparrow},\ket{\uparrow\uparrow}$.
The eigenvalues of Hamiltonian \eqref{eq:exact Hamiltonian} in this subspace are
\begin{align}
    E_\pm=\Delta\pm\frac\Omega{\sqrt2},&& E_{0}=\Delta.
    \label{eq:eigenenergy two facilitated atoms}
\end{align}
The eigenstate of corresponding to eigenvalue $E_0$ is the antisymmetric superposition $(\ket{\uparrow\downarrow}-\ket{\downarrow\uparrow})/{\sqrt 2}$.
Since the operator that effectuates transitions is conserving the symmetry of the initial state $\ket{\downarrow\downarrow}$, see Eq.~\eqref{eq:Ham spectroscopy}, the transition rate to the antisymmetric state vanishes and it cannot be excited in the spectroscopy. Hence, only two spectral lines are observed. 
In \figref{fig:time-resolved spectroscopy}{(a)} we present the time evolution of the Rydberg density $\ev{n}_t=\ev{n_1}_t+\ev{n_2}_t$ as a function of the modulation frequency $\omega_\mathrm{s}/V_1$.
As can be seen, the two expected spectral lines become visible only when the time exceeds $\Omega t>5$.
This shows that the spectroscopy requires a minimal coherence time in order to reach the necessary resolution.
In \figureref{fig:time-resolved spectroscopy}{(b,c)} we show the time-averaged Rydberg density $\bar n=\frac{1}{\tau}\int_{0}^{\tau}\mathrm dt \ev{n}_t$ with an integration time $\tau_1 = 2/\Omega$, panel~\textbf{(b)}, and $\tau_2=10/\Omega$, panel~\textbf{(c)}. 
\bibliography{references}


\end{document}